\documentclass[conference]{IEEEtran}

\usepackage{cite}

\ifCLASSINFOpdf
 \usepackage[pdftex]{graphicx}

\else

\fi

\begin{document}
%
\title{Work Motivational Challenges Regarding the Interface Between Agile Teams and a Non-Agile Surrounding Organization: A case study}


\author{\IEEEauthorblockN{Lucas Gren}
\IEEEauthorblockA{Chalmers University and \\Gothenburg University\\
Gothenburg, Sweden 412--92\\
Email: lucas.gren@cse.gu.se}
\and
\IEEEauthorblockN{Richard Torkar}
\IEEEauthorblockA{Chalmers University and \\Gothenburg University\\
Gothenburg, Sweden 412--92 and\\
Blekinge Institute of Technology\\
Karlskrona, Sweden 371--79\\
Email: richard.torkar@cse.gu.se}
\and
\IEEEauthorblockN{Robert Feldt}
\IEEEauthorblockA{Blekinge Institute of Technology\\
Karlskrona, Sweden 371--79\\
Email: robert.feldt@bth.se}}

\maketitle

\begin{abstract}
There are studies showing what happens if agile teams are introduced into a non-agile organization, e.g. higher overhead costs and the necessity of an understanding of agile methods even outside the teams. This case study shows an example of work motivational aspects that might surface when an agile team exists in the middle of a more traditional structure. This case study was conducted at a car manufacturer in Sweden, consisting of an unstructured interview with the Scrum Master and a semi-structured focus group. The results show that the teams felt that the feedback from the surrounding organization was unsynchronized resulting in them not feeling appreciated when delivering their work. Moreover, they felt frustrated when working on non-agile teams after have been working on agile ones. This study concludes that there were work motivational affects of fitting an agile team into a non-agile surrounding organization, and therefore this might also be true for other organizations.
\end{abstract}

\IEEEpeerreviewmaketitle

\section{Introduction}
There are many success stories of companies that have transitioned to an agile way of working. In complex projects where a clear goal and finish line are hard to define and ever-changing, a more flexible managerial style is often needed \cite{engwall}. With increasing success in ``saving'' projects in crisis and with these projects being of different sizes and having diverse circumstances there was a more widespread acceptance and use of agile approaches to software development. Over time the concept also saw increasing use as a more general approach to project management. Agile thinking and methods are not an isolated phenomenon though. According to \cite{fernandez} several books covering different management schemes and theories have been written that relates to and touches on the ideas underlying agile project management practices, such as: \emph{Critical Chain Theory} \cite{goldratt} and \emph{Lean Production} \cite{feld}. 

The benefits of introducing agile methods in organization have been proven to be mostly positive for many organizations, e.g.\ \cite{laanti}. It has also been shown that job satisfaction increase on agile teams \cite{melnik2}. However, motivational aspects of agile teams' interface to a surrounding non-agile organization have not been found. This study aims to show an example of what could happen to agile team members' motivation when working on an agile team in a larger non-agile environment. Therefore, the research question is ``is there job motivational aspects regarding the interface between agile teams and a non-agile surrounding organization?''.

\section{Large Organizations and Agile Methods}
\subsection{Traditional Project Management}
In traditional project management, trade-offs are often made between time, cost, and quality, i.e.\ it is impossible to prioritize all three \cite{turner2, babu, khang}. In order to choose from different projects many organizations select according to a set of financial decision methods. These are often simply based on cash flows (which has evident drawbacks; for example, how to put monetary value on other resources \cite{granstrand}), but are widely used.

\paragraph{Net Present Value}
The most common method is the Net Present Value (NPV) approach. This method is based on the assumption that money is worth more today than in the future (the time value of money). This means that future earnings are worth less today so its value reflects a discount. Therefore, this rate is referred to as a discount rate $r$. This means that the sum of all cash flows discounted for today is the present value of a project: $\mathrm{PV}=\sum_{n=0}^{N} \frac{C_{n}}{(1+r)^{n}}$. Where $C_{n}$ is future value for the investment at year $n$ (and $C_{0}$ is present day). All the present values of all the cost and earnings for the project is thereby calculated. The Net Present Value is then: $\mathrm{NPV} = \mathrm{PV}(\mathrm{benefits}) - \mathrm{PV}(\mathrm{costs})$. This means that if the $\mathrm{NPV}>0$ the project is worth running. An interesting fact is also that a firm's total value is the NPV of all assets in it \cite{berk}. One critique against the NPV approach is that it assumes only one decision point in the beginning of the project. A part of a solution could be a Real Options Approach. In this case, different paths and cash flows are calculated and weighted according to their probabilities \cite{berk, copeland}. This is also connected to the stage-gate methods described below. 

The above approach is considered to be a good way to get an overview of project activities and is also fairly simple to draw and understand. The simplicity will bring disadvantages such as the difficulty to make updates when many changes are needed, and they do not help in optimizing resource allocation. That in itself would imply that they could bring a false sense of certainty about the project, often connected to time estimation.

Time estimation is affected by many different factors. One of them is learning effects that can be described with a learning curve: $Y_{x} = Kx^{n}$ where $x$ is the number of times the task has been carried out, $Y_{x}$ is the time taken to carry out the task the $x$th time, $K$ is the time it took the first time, and $n$ is $ \frac{\ln b}{\ln 2}$ where $b$ is the learning rate \cite{maylor, granstrand}. In addition, one intimate factor to consider when performing time estimation is risk.

There is a basic stage-gate system for splitting a project into parts (or stages), as defined by \cite{cooper}. This way a project must pass through a gate before proceeding to the next stage. The purpose is to solve problems where they pop up and not to pass them on to the next stage. A disadvantage is that a new part of the process cannot start before the previous one is done. Concurrent engineering, i.e.\ to make the processes overlap, can solve this partially. The basic idea when planning a project would then be to break it down into small tasks. This is often done through a work breakdown structure. The second step is to construct a time plan according to this structure in order to estimate how long time the project will take.

There is a set of techniques to conduct Risk Management, but they should all include identification, quantification and mitigation. In order to quantify the risks one can assess the severity, hide-ability, and likelihood of a risk in order to get a Risk Priority Number (by simply multiplying these scores). Other, more quantitative, methods are expected value, Monte Carlo simulation, and PERT. The latter is a simple way to add optimistic and pessimistic times to all estimates in order to more accurately get a time approximation that is not based on a single guess. This adds probability to the time estimation process, which will, hopefully, make it more realistic \cite{hartley}.

\subsection{Agile Project Management}
The basic idea of agile project management is that complex projects need to combine the traditional approach to managing projects and the need to be able to respond to change. The agile community has, thus, defined a set of principles that they summarize in The Agile Manifesto \cite{fowler2001}:

\begin{itemize}
\item Individuals and interactions over processes and tools.
\item Working software over comprehensive documentation.
\item Customer collaboration over contract negotiation.
\item Responding to change over following a plan. 
\end{itemize}

Many customers have business needs that change over time, reflecting not only new needs but also the need to respond to a change in the marketplace. There are many agile practices such as eXtreme Programming (XP), Crystal, and Scrum, which try to take this into account. In Scrum the project has a prioritized backlog of requirements and use iterative development (called `sprints') to get basic working software for the customer to view as soon as possible. Scrum uses self-organizing teams that get coordinated through daily meetings called `scrums'. Agile development, in general, is customer-focused, which means that the customer is preferably on site. This means that the project is not strictly planned up front, but changes continuously throughout the project. Instead of having activities planned exactly the project maintains a flexibility that is needed in order to rapidly respond to change. The managerial culture of agile methods is trust, commitment, teamwork, equality, and fair treatment. This means that agile methods will probably work best in flat organizations and have aligned decision-making on all levels \cite{moe2012}. Further, agility must be present at all levels including the strategic one \cite{roth}. The idea is to have evidence-based decisions, goal-focus (with change built in), independence with responsibility, and long-term thinking also known as sustainable pace (i.e.\ a 40-hour workweek). The manager of an agile team tries to generate group effectiveness by being a facilitator and not a supervisor, and transparency is key for this process to work \cite{schwaber, chin}.

\subsection{Agility and Discipline}
In software engineering the traditional approach to software development projects is usually considered to be `Plan-Driven'. These methods come from the systems engineering and other disciplines, and were established to coordinate large inter-operating components. Software does not function as hardware and, therefore, different standards were introduced. The basic assumption is that software engineering is a process of formal mathematical specification and verification. The process is divided into different steps (i.e.\ a waterfall), which are thoroughly documented. The process is standardized, and incrementally improved to control and manage the work-flow \cite{boehmm}. When changing to an agile method, where cooperation and self-organizing team are central, some aspects of the modern workplace might cause problems. If group members are unable to, e.g., be physically present during meeting, the aspect of human interaction becomes harder to achieve and problems concerning communication, culture, trust, and knowledge management appear \cite{jala}. There are also some indications that people that does not have programming responsibilities in a large organization think that agile methods is unsuitable in general \cite{korhonen}. There is also an aspect of integrating flexibility in fixed and large organizations. Agile methods can give a traditional stage-gate model a powerful micro-planning tool and increase the change response time. If the whole organization has not embraced the agile principles, an agile team that adapts to a stage-gate system can synchronize their development with other teams and functions of the organization. In order to make this feasible, the agile team must be prepared to interface with the traditional stage-gate system around it. The important part is that the team is aware of these extra overhead costs. Agile methods are generally more accepted by team members and more feared by management. However, in order to make this work a universal acceptance in the organization is much needed \cite{karlstrom}. However, it has been some evidence showing that agile teams have higher job motivation than non-agile teams \cite{melnik2}.

\section{Method}
The methodology used for this study consisted of an interview with the Scrum Master and a focus group with two teams participating. 

\subsection{Case and Subjects Selection}\label{sub:case_and_subject_selection}
The teams in this study were two teams with the same Scrum Master at Company X in Sweden. Company X is a part of a larger firm, which provides world-wide supply chain expertise to a set of automotive companies. The IT part is, of course, essential for the company to function. Many organizations, independent of field, need an efficient IT department to provide good solutions for the whole organization. The organization took a decision to implement agile methods and was conducting a first pilot study to later diffuse the methods to more parts of the organization. 

The teams that were a part of this study had the task of developing an extension of a corporate software system used for supply chain management. In their work process they integrated agile methods and Scrum specifically. The reason why this case is from software engineering is that they have the most experience with agile methods and were easier to find. This software project included many teams, but two of these teams were using Scrum and had the same Scrum Master. The groups were a mix of business and programming focused employees and external resources. The reason for this mix was to assert that the business aspects of the project were considered and to create a method that more areas in the organization could use. Many of the team members had therefore management tasks. Since there were unclear separation between the two teams and the fact that they had the same Scrum Master we chose to meet both teams collectively.

\subsection{Data Collection Procedures}\label{sub:data_collection_procedures}
The first contact with the company was via an unstructured 40-minute interview the Scrum Master of these new agile projects. During the interview one researcher were taking notes carefully. The Scrum Master then set up a meting inviting all members from both teams ($N=23$). A subset of these team members attended the meeting/focus group ($N=10$). The team members were informed that they would evaluate their new process in a focus group with a researcher from university. We had a set of questions to start the discussion (semi-structured group interview\slash focus group), however, the team had a lot to say about their new ways of working and its connection to the rest of the organization. The topics covered were:

\begin{itemize}
\item The teams' experience with\slash opinions of their new agile process.
\item A comparison with their other current projects.
\item Differences between this project and others they have experienced.
\end{itemize}

One researcher participated during the one-hour focus group and carefully wrote down what being said.  The interviews were not recorded since we wanted participant to be able to speak as freely as possible regarding their emotions connected to their participation on the team. The tradeoff is then, of course, that we cannot say exactly how many times each individual agreed on a topic lifted by one of their colleague. The researcher who participated in the focus group wrote down aspects the team focused on during the session instead.

\subsection{Analysis Procedures}\label{sub:analysis_procedures}
After both the interview with the Scrum Master and the focus group the notes were carefully reviewed and summarized by one author. The summaries were thematically analyzed only keeping statements regarding work motivation. After this, the statements were categorized, and compared to other research. For example, unsynchronized feedback loops were mentioned by several individuals and no other participants expressed disagreement. Therefore this aspect was interpreted as important and presented below.

\section{Findings}
\subsection{Summary of Interview with the Scrum Master}
The Scrum Master of the two teams describes the system they are developing and the first enterprise system project so far for them. The purpose is to integrate this new system into the rest of the organization and the system is safety-critical. The organization traditionally has a stage-gate project management method that is very strict. This framework is fixed and they have to adapt to it and deliver what is needed at certain milestones. Both these milestones and a budget for the whole project must be predefined. The idea with agile methodology is to work agile in between the gates at different stages. They use a plug-in iteration process of agile that is not exactly what they expressed that they wanted in the beginning of the project. The business part of the project had been going on for half-a-year already, and they have two-week sprints with systems specifications to each sprint. The total amount of sprints is nine, and they have a meeting at day five in every sprint. The project uses a more strict way of writing requirements and they do not apply user stories. The get their requirements from the product owner, and this person decides on the requirements and their priority. The get the requirements documents to the teams by a standard called ``Business Rules Description''. The got a prototype up and running fast with basic functionality.

\subsection{Summary from the focus group}
Some members expressed stress connected to the feedback system from the surrounding part of the organization. If they had struggled to reach a deadline internally within the group, the effort was not recognized by other parts of the organization since they had other milestones to follow. They would have preferred to stop and celebrate somewhat and then move on. At other times, they received positive feedback from managers without them being even close to a delivery. This was described and odd and unsynchronized. Then the members had a discussion about how working agile had helped them in their group development process. The Scrum practices had given them a forum and a place to discuss solutions and conflicts on a regular basis before they become more infected. The members who were not 100\% dedicated to the team but had other concurring projects said they really felt a difference between the two. The other projects felt slow and unresponsive and they had gotten used to rapid responses and quick progress with issues. They all agreed that job satisfaction was higher for them when working on the agile team. They also compared their result to another non-agile team, and stated that they were way ahead of them considering what they had delivered.

\section{Discussion}
To use agile methods instead of traditional project management has its advantages \cite{schwaber, petersen2}. It was also mentioned by the focus group that the agile work group had better results than other groups within the company. However, they had to create a project plan and a budget before the project started, and had to adapt to the organization's surrounding stage-gate project management tools. There are often problems when trying to scale up or use agile in a larger context \cite{petersenhej, boehmm} and the whole organization must accept and know the difference in how the agile team is working \cite{karlstrom}. 

One of the advantages with agile development compared to other traditional methods is that decisions about the final product can be made underway. One of the critiques of the Net Present Value described is that it only has one decision point in time \cite{granstrand}. Agile methods make decisions possible underway. This can also be argued as being more honest, since this mostly happens anyway but often shadowed by cover-up explanations \cite{engwall}. The aspect of stakeholder analysis is also different in agile project management where the customer has to make new decisions about the product underway. Since the product owner in the studied organization owned the requirements and sent them to the teams, the stakeholder analysis seemed to still be traditional and not as much part of the team as suggested in agile development. 

The studied groups probably would get more positive effects out of working agile if the organization around it would not have been maintaining a stage-gate system. Taking a budget decision before the project has started, which then is not possible to change, locks the project into a certain way of resisting flexibility in order to deliver what was expected from the beginning. The Scrum Master also described this as a problem, since the end-cost had to be decided beforehand. This seems to be a dilemma when this large company tries to implement flexibility, but does not dare to be flexible about budgets and goals to conduct a project. The problem is then that, if the project is very complex and no final goal can be decided with almost any certainty, the final result will not be as good as it could have been. \cite{engwall} even states that the dream of the perfect goal is futile, and the organization probably lacks the trust needed to let go of some control. 

However, there are reasons for companies not to implement agile fully in all aspects of a project. Aspects such as, that plans drive funding, different people do architecture and design, documents are needed to mitigate risks, and development is not a part of the requirements process are all reasons to maintain some traditional methods \cite{west}. All of these were reasons why the studied teams did not fully implement the agile concepts. However, the main contribution of this study is that in order to combine agile teams with a stage-gate/traditional project organization the organization surrounding the teams must understand how the agile team is different and adapt their feedback to their way of working. If they do not, the agile team's earlier increased job motivation will somewhat decrease. An agile team, just like any other team, expects feedback when they have worked hard and delivered a good result. As this case study shows, the motivation of the team will, of course, decrease if the surrounding organization lacks the understanding of the agile team's different ways of working. Furthermore, if the positive or negative feedback is given to the team at the wrong time, the team will feel that their efforts are not appreciated. Since most organizations combine agile with traditional projects management this study shows that they should be aware of the interface between the agile teams and the surrounding organization, not just for overhead costs reasons, but also from a work motivational perspective. In order to mitigate this lowered work motivation risk, companies could make sure the stages and gates of the surrounding organization are, at least somewhat, synchronized with the iterations of the agile teams. This way, the most evident feedback disappointments could be avoided. 

In this specific case study the teams seemed to be content with their methodology and made comparisons with other non-agile teams in the same department. Compared to them, the agile projects had delivered more value and faster, which confirms earlier success stories from agile software development \cite{schwaber}. The focus group result shows that team members were more motivated on the agile teams than when they were working on other teams. The confirms job satisfaction and team spirit research already conducted by for example \cite{melnik2}. This study adds the perspective of employees getting frustrated when working on non-agile teams after being on an agile one within the same organization, due to a different pace. This problem is harder to address, but making the employees aware of these effects beforehand might decrease their disappointment and frustration.

\section{Limitations}
This case study only shows one example of what happened to agile teams in a larger non-agile organization. In connection to research of what motivates employees, it is most likely that unsynchronized feedback loops will have the same effect in other organizations. However, we, of course, cannot conclude that the problem of unsynchronized feedback loops or frustration when returning to non-agile teams, are occurring in other organizations, since we only studies one. Therefore, this study only presents what happened on these specific teams. Furthermore, the methodology is not thorough. It would have been a good idea to check the reliability of the thematic analysis by having more researchers code what was said in the interview and the focus group. 

\section{Conclusion and Future Work}
In conclusion, this study has shown that challenges when integrating agile teams into a surrounding non-agile organization, did not only regard overhead costs, but were also of job motivational nature in this specific case. This was due to unsynchronized feedback loops between the agile teams' delivery points and the surrounding stage-gate milestones. Furthermore, employees reported being frustrated with the slow pace when working on a non-agile team after having been on an agile team. 

These issues might occur on other organizations as well, and if it does, it is most likely that the agile teams will feel unappreciated since e.g. positive feedback will not be given to the teams when they expect it. This result was shown as a case study including an interview with the Scrum Master and a focus group with a subset of two agile teams. 

This means that organizations that implement agile methods within traditional organizations should, not only expect higher overhead costs, but also be aware of the different feedback loops needed to the agile teams, and expect lower motivation and frustration from employees on non-agile teams after have participated in an agile project. 

This study was just a first step in studying work motivational aspects regarding the interface between agile teams and a non-agile surrounding organization. The most obvious future work is to see if these findings are true in more organizations, which is preferably investigated using both qualitative (e.g.\ by conducting more interviews using a more thorough method) and quantitative data (e.g.\ distributing a survey to see how agile teams in large non-agile organizations end up on work motivation scales connected to this subject).

\section*{Acknowledgment}
We would like to thank Pasi Moisander, Karin Scholes, and Kristin Boissonneau Gren (without your goodwill this work could not have been done).

\bibliographystyle{IEEEtran}

\bibliography{references}

\begin{thebibliography}{10}
\providecommand{\url}[1]{#1}
\csname url@samestyle\endcsname
\providecommand{\newblock}{\relax}
\providecommand{\bibinfo}[2]{#2}
\providecommand{\BIBentrySTDinterwordspacing}{\spaceskip=0pt\relax}
\providecommand{\BIBentryALTinterwordstretchfactor}{4}
\providecommand{\BIBentryALTinterwordspacing}{\spaceskip=\fontdimen2\font plus
\BIBentryALTinterwordstretchfactor\fontdimen3\font minus
  \fontdimen4\font\relax}
\providecommand{\BIBforeignlanguage}[2]{{%
\expandafter\ifx\csname l@#1\endcsname\relax
\typeout{** WARNING: IEEEtran.bst: No hyphenation pattern has been}%
\typeout{** loaded for the language `#1'. Using the pattern for}%
\typeout{** the default language instead.}%
\else
\language=\csname l@#1\endcsname
\fi
#2}}
\providecommand{\BIBdecl}{\relax}
\BIBdecl

\bibitem{engwall}
M.~Engwall, ``The futile dream of the perfect goal,'' \emph{Beyond project
  management}, vol.~1, pp. 261--277, 2002.

\bibitem{fernandez}
D.~Fernandez and J.~Fernandez, ``Agile project management--{A}gilism versus
  traditional approaches,'' \emph{Journal of Computer Information Systems},
  vol.~49, no.~2, pp. 10--17, 2008.

\bibitem{goldratt}
E.~Goldratt, \emph{Critical chain}.\hskip 1em plus 0.5em minus 0.4em\relax
  Great Barrington: North River Press, 1997.

\bibitem{feld}
W.~Feld, \emph{Lean manufacturing: {T}ools, techniques, and how to use
  them}.\hskip 1em plus 0.5em minus 0.4em\relax Boca Raton, Fla.: St. Lucie
  Press, 2001.

\bibitem{laanti}
M.~Laanti, O.~Salo, and P.~Abrahamsson, ``Agile methods rapidly replacing
  traditional methods at {N}okia: {A} survey of opinions on agile
  transformation,'' \emph{Information and Software Technology}, vol.~53, no.~3,
  pp. 276--290, 2011.

\bibitem{melnik2}
G.~Melnik and F.~Maurer, ``Comparative analysis of job satisfaction in agile
  and non-agile software development teams,'' in \emph{Extreme Programming and
  Agile Processes in Software Engineering}.\hskip 1em plus 0.5em minus
  0.4em\relax Springer, 2006, pp. 32--42.

\bibitem{turner2}
R.~Turner, \emph{The handbook of project-based management: improving the
  processes for achieving strategic objectives}, 2nd~ed.\hskip 1em plus 0.5em
  minus 0.4em\relax London: McGraw-Hill, 1999.

\bibitem{babu}
A.~Babu and N.~Suresh, ``Project management with time, cost, and quality
  considerations,'' \emph{European Journal of Operational Research}, vol.~88,
  no.~2, pp. 320--327, 1996.

\bibitem{khang}
D.~Khang and Y.~Myint, ``Time, cost and quality trade-off in project
  management: a case study,'' \emph{International journal of project
  management}, vol.~17, no.~4, pp. 249--256, 1999.

\bibitem{granstrand}
O.~Granstrand, \emph{Industrial innovation economics and intellectual
  property}, 5th~ed.\hskip 1em plus 0.5em minus 0.4em\relax G{\"o}teborg:
  Svenska kulturkompaniet, 2010.

\bibitem{berk}
J.~Berk and P.~DeMarzo, \emph{Corporate finance}, 2nd~ed.\hskip 1em plus 0.5em
  minus 0.4em\relax Harlow, Essex: Pearson, 2011.

\bibitem{copeland}
T.~Copeland, F.~Weston, and K.~Shastri, \emph{Financial theory and corporate
  policy}, 4th~ed.\hskip 1em plus 0.5em minus 0.4em\relax Boston, Mass.:
  Pearson Addison-Wesley, 2005.

\bibitem{maylor}
H.~Maylor, \emph{Project management}, 4th~ed.\hskip 1em plus 0.5em minus
  0.4em\relax Harlow, England: Financial Times Prentice Hall, 2010.

\bibitem{cooper}
R.~Cooper, ``Stage-gate systems: a new tool for managing new products,''
  \emph{Business Horizons}, vol.~33, no.~3, pp. 44--54, 1990.

\bibitem{hartley}
H.~Hartley and A.~Wortham, ``A statistical theory for pert critical path
  analysis,'' \emph{Management Science}, vol.~12, no.~10, pp. B--469, 1966.

\bibitem{fowler2001}
M.~Fowler and J.~Highsmith, ``The agile manifesto,'' In Software Development,
  Issue on Agile Methodologies, last accessed on December 29th, 2006, Aug.
  2001.

\bibitem{moe2012}
N.~Moe, A.~Aurum, and T.~Dyb{\aa}, ``Challenges of shared decision-making: A
  multiple case study of agile software development,'' \emph{Information and
  Software Technology}, vol.~54, no.~8, pp. 853--865, 2012.

\bibitem{roth}
A.~Roth, ``Achieving strategic agility through economies of knowledge,''
  \emph{Strategy \& leadership}, vol.~24, no.~2, pp. 30--36, 1996.

\bibitem{schwaber}
K.~Schwaber, \emph{Agile project management with Scrum}.\hskip 1em plus 0.5em
  minus 0.4em\relax Redmond, Wash.: Microsoft Press, 2004.

\bibitem{chin}
G.~Chin, \emph{Agile project management: how to succeed in the face of changing
  project requirements}.\hskip 1em plus 0.5em minus 0.4em\relax New York:
  AMACOM, 2004.

\bibitem{boehmm}
B.~Boehm and R.~Turner, ``Management challenges to implementing agile processes
  in traditional development organizations,'' \emph{Software, IEEE}, vol.~22,
  no.~5, pp. 30--39, 2005.

\bibitem{jala}
S.~Jalali and C.~Wohlin, ``Agile practices in global software engineering--a
  systematic map,'' in \emph{International Conference on Global Software
  Engineering (ICGSE)}.\hskip 1em plus 0.5em minus 0.4em\relax IEEE, 2010.

\bibitem{korhonen}
K.~Korhonen, ``Adopting agile practices in teams with no direct programming
  responsibility--{A} case study,'' \emph{Product-Focused Software Process
  Improvement}, pp. 30--43, 2011.

\bibitem{karlstrom}
D.~Karlstr{\"o}m and P.~Runeson, ``Combining agile methods with stage-gate
  project management,'' \emph{Software, IEEE}, vol.~22, no.~3, pp. 43--49,
  2005.

\bibitem{petersen2}
K.~Petersen and C.~Wohlin, ``The effect of moving from a plan-driven to an
  incremental software development approach with agile practices,''
  \emph{Empirical Software Engineering}, vol.~15, no.~6, pp. 654--693, 2010.

\bibitem{petersenhej}
------, ``A comparison of issues and advantages in agile and incremental
  development between state of the art and an industrial case,'' \emph{Journal
  of Systems and Software}, vol.~82, no.~9, pp. 1479--1490, 2009.

\bibitem{west}
D.~West \emph{et~al.}, ``Water-scrum-fall is the reality of agile for most
  organizations today,'' \emph{Forrester Research, July}, vol.~26, 2011.

\end{thebibliography}

\end{document}